\documentclass[10pt,journal,twocolumn]{IEEEtran}
\IEEEoverridecommandlockouts
\usepackage{algpseudocode}
\usepackage{algorithm}
\usepackage{bm}
\usepackage{amsfonts}
\usepackage{amssymb}
\usepackage{amscd}
\usepackage{graphics}
\usepackage[dvips]{graphicx}
\usepackage{epsfig}
\usepackage{subfigure}
\usepackage[cmex10]{amsmath}
\usepackage{array}
\usepackage{eqparbox}
\usepackage{flushend}
\usepackage{hyperref}
\usepackage{fancyhdr}
\usepackage{titling}
\usepackage{enumerate}

\DeclareMathAlphabet{\mathbsf}{OT1}{cmss}{bx}{n}
\DeclareMathAlphabet{\mathssf}{OT1}{cmss}{m}{sl}
\DeclareMathAlphabet{\mathcsf}{OT1}{cmss}{sbc}{n}

\newcommand{\ie}{{\em i.e.}}
\newcommand{\etc}{{\em etc}}
\newcommand{\eg}{{\em e.g.}}

\newcommand{\secref}[1]{Section~\ref{#1}}
\newcommand{\figref}[1]{Fig.~\ref{#1}}
\newcommand{\tabref}[1]{Table~\ref{#1}}

\makeatletter
\def\blfootnote{\xdef\@thefnmark{}\@footnotetext}
\makeatother

\renewcommand{\baselinestretch}{1}
\newcommand{\qed}{\nobreak \ifvmode \relax \else
      \ifdim\lastskip<1.5em \hskip-\lastskip
      \hskip1.5em plus0em minus0.5em \fi \nobreak
      \vrule height0.75em width0.5em depth0.25em\fi}

\def\BibTeX{{\rm B\kern-.05em{\sc i\kern-.025em b}\kern-.08em
    T\kern-.1667em\lower.7ex\hbox{E}\kern-.125emX}}

\usepackage{newlfont}

\begin{document}
\title{A Unifying Framework for the Electrical Structure-Based Approach to PMU Placement in Electric Power Systems}
\author{\IEEEauthorblockN{K. G. Nagananda and Shalinee Kishore}\thanks{K. G. Nagananda was with the Department of Electrical and Computer Engineering, Lehigh University, Bethlehem, PA $18015$, U.S.A. E-mail: \texttt{kgn209@lehigh.edu}. Shalinee Kishore is with the Department of Electrical and Computer Engineering, Lehigh University, Bethlehem, PA $18015$, U.S.A. E-mail: \texttt{skishore@lehigh.edu}}
}

\pagenumbering{gobble}
\date{}
\maketitle

\begin{abstract}
The electrical structure of the power grid is utilized to address the phasor measurement unit (PMU) placement problem. First, we derive the connectivity matrix of the network using the resistance distance metric and employ it in the linear program formulation to obtain the optimal number of PMUs, for complete network observability without zero injection measurements. This approach was developed by the author in an earlier work, but the solution methodology to address the location problem did not fully utilize the electrical properties of the network, resulting in an ambiguity. In this paper, we settle this issue by exploiting the coupling structure of the grid derived using the singular value decomposition (SVD)-based analysis of the resistance distance matrix to solve the location problem. Our study, which is based on recent advances in complex networks that promote the electrical structure of the grid over its topological structure and the SVD analysis which throws light on the electrical coupling of the network, results in a unified framework for the electrical structure-based PMU placement. The proposed method is tested on IEEE bus systems, and the results uncover intriguing connections between the singular vectors and average resistance distance between buses in the network.
\end{abstract}


\section{Introduction}\label{sec:introduction}
\subsection{Problem and solution methodology}\label{subsec:problem_methodology}
In the classic setting, the phasor measurement unit (PMU) placement problem in electric power systems is divided into two parts: (a) the task of obtaining the optimal or minimum number of PMUs and (b) to find the optimal locations to install these PMUs on the power grid to meet a desired objective, which is usually complete or incomplete network observability and with or without zero injection measurements. In this paper, we provide a complete picture of the relevance of the electrical structure of the power network to the above stated parts of the PMU placement problem. Specifically, we first derive the adjacency matrix of the grid using the resistance distance metric \cite{Klein1993} rather than direct topological connections between buses in the grid, and employ it in the linear program formulation to obtain the optimal set of PMUs. Next, for part (b), we exploit the electrical coupling structure of the network obtained using the singular value decomposition (SVD) of the resistance distance matrix derived to address part (a). In this paper, we restrict our analysis to the case of complete network observability, and without zero injection measurements. 

\subsection{Main contribution}\label{subsec:main_contribution}
The electrical structure-based approach to obtain the minimum number of PMUs was first devised by the author in \cite{Nagananda2013a}, where the optimal location problem was addressed by introducing the notion of average resistance distance in conjunction with the graphical structure of the adjacency matrix (see \cite[Sections IVA, IVB]{Nagananda2013a}). However, the solution methodology developed there did not fully exploit the electrical properties of the network, leading to an ambiguity in placing the optimal set of PMUs, which was derived by solving the integer program. In this paper, we settle the issue that arose in \cite{Nagananda2013a} by exploiting the coupling structure of the grid to solve the location problem. Therefore, the procedure developed in \cite[Section IIIB]{Nagananda2013a} to obtain the optimal set of PMUs together with the SVD-based analysis developed in this paper to solve the location problem results in a unified framework for the electrical structure-based PMU placement in electric power systems. In the process, we uncover intriguing connections between the singular vectors of the distance matrix and the average resistance distance between various buses in the network.

The method proposed in \cite{Nagananda2013a} to solve the location problem and the resulting ambiguity is presented in \secref{subsec:location_ambiguity}. We begin by justifying the use of the electrical structure of the grid over its topological structure for PMU placement.
\vspace{-0.25cm}
\subsection{Electrical structure versus topological structure}\label{subsec:topo_vs_elec}
The electric power grid has received considerable attention from the perspective of complex networks \cite{Dorfler2010}. In the following we briefly present this perspective, which promotes the electrical structure of the grid over its topological structure.

In \cite{Cotilla-Sanchez2012} (see Section I and references therein), it was reported that electric grids in different geographical locations had different degree distributions leading to varied topological structures. It was also pointed out that the same grid had different topological structures by carrying out different model-based analyses. This discrepancy was attributed to the weaker characterization of the electrical connections between network components as provided by the topological structure. Related reports supporting this line of argument were found in \cite{Wu1995} - \nocite{Wu2005}\cite{Atay2006}, where it was shown that, for many classes of complex networks, characterizing the network structure using degree distribution alone was suboptimal and had implications on node synchronization and performance of the network.

In the context of PMU placement, for the topology-based approach, the bus admittance matrix plays a central role in solving the placement problem. Though the admittance matrix characterizes the electrical behavior of the network, the sensitivity between power injections and nodal phase angle differences can be utilized to better characterize the electrical influence between network components. The first step in this direction is to measure the amount of electrical influence between different components in the network, and is summarized in the following paragraph.

The measurement of the electrical influence necessitates a metric system. This can be devised by deriving the sensitivity matrix and taking its complement to obtain the distance matrix, whose entries quantify the electrical influence that each component has on the other - zero value indicates that two components are perfectly connected, while a large number indicates that the corresponding components have negligible electrical influence on each other. The electrical distance was proved to be a formal distance metric, and was employed to address various problems in power systems (see \cite[Section III]{Cotilla-Sanchez2012} and the references therein). The resistance distance, employed in this paper, is one such metric which provides a strong characterization of the electrical connectedness between nodes in the network. Likewise, the electrical coupling measures the connectivity between nodes in the network, and is obtained by computing the magnitude of the entries of the singular vectors derived from the SVD of the resistance distance matrix.

Following these advances in the metric system, the electrical structure-based approach provides a more comprehensive characterization of the electrical connectivity between buses in the grid, and hence is more favorable than the topology based approach to address the PMU placement problem.

The remainder of the paper is organized as follows. In \secref{sec:background}, we present a summary of related previous work on PMU placement and the SVD analysis for power systems. In \secref{sec:optimalset_pmu}, we formulate the integer linear program to obtain the optimal set of PMUs for both the topological and electrical structures based approaches. In \secref{sec:pmu_placement}, we perform the SVD of the bus admittance and resistance distance matrices to derive the coupling structure used for the optimal placement of these PMUs. Simulation results and related discussion are provided in \secref{sec:results_discussion}. We conclude the paper in \secref{sec:conclusion}.
\vspace{-0.10cm}
\section{Literature review}\label{sec:background}
In this section, we first summarize previously published work on PMU placement, and then compare our work with two closely related papers that employ similar SVD-based analysis.
\vspace{-0.6cm}
\subsection{Work on PMU placement}\label{subsec:literature_pmuplacement}
The PMU placement problem is formulated in the mathematical programming framework for complete and incomplete network observability \cite{Nuqui2005}, with and without zero power injection measurements. In \cite{Baldwin1993}, the minimal PMU set was obtained using a dual search bisecting simulated-annealing algorithm searches for complete network observability, while the location problem was solved using a spanning measurement subgraph. A non-dominated sorting genetic algorithm yielding a Pareto-optimal solution was presented in \cite{Milosevic2003}, where the integer program exhibited nonlinearity in the presence of power injection measurements. An integer program was formulated in \cite{Xu2004} to include conventional power flow and injection measurements in addition to PMU measurements for maximum network observability, while in \cite{Chen2006}, a strategy was devised for bad data detection during state estimation using optimal PMU placement. The placement problem was formulated as an integer linear program for complete and incomplete network observability, with and without conventional power flow and injection measurements in \cite{Gou2008},\cite{Gou2008a}.

Multiple placement solutions was proposed in \cite{Zhang2010}, in the framework of power system dynamic state estimation. The affinity propagation algorithm to solve the integer linear program was proposed in \cite{Deka2011}, which also addressed correction of measurement errors in the PMUs. In \cite{Azizi2012}, state estimation using phasor measurements with complete network observability was shown to be linear, where an exhaustive search-based method was devised to solve the placement problem. An estimation-theoretic criteria to optimize PMU placement was considered in \cite{Kekatos2012}, where the problem was solved using a convex relaxation method incorporating system states estimated within a Bayesian framework. In \cite{Li2013}, an information-theoretic measure, namely, mutual information (MI) was employed to address the PMU placement problem, where the objective was to maximize the MI between PMU measurements and power system states to obtain highly ``informative'' PMU configurations; a counterargument to the analysis presented in this work was reported in \cite[Section IVC]{Nagananda2013a}.

However, studies in the above references do not fully utilize the electrical influence between components in the power grid. Therefore, though the results are interesting in their own right, they do not provide a realistic scenario for optimal system operation, thus motivating the study in this paper.
\vspace{-0.33cm}
\subsection{Comparing our work with \cite{Wang2010} and \cite{Dehghani2013}}\label{subsec:literature_svd}
In this subsection, we begin with references \cite{Wang2010} and \cite{Dehghani2013} that are closely related to the work presented in this paper. In \cite{Wang2010}, the coupling structure of the grid was explored by performing the SVD of the network admittance matrix, where it was shown that the unitary matrices comprising the singular vectors are sparse. This SVD sparsity was exploited to perform power flow calculations on a reduced or compressed network leading to robust decentralized control and communication architectures. However, the goal there was to compress the electrical grid, and there was no instance of PMU placement in the exposition. In \cite{Dehghani2013}, the SVD was employed to solve the PMU placement problem for complete network observability to achieve dynamic stability assessment in the network. The SVD of Prony matrix was used to rank the PMUs according to their effect on dynamic stability. Our work is different compared with the aforementioned references in that our main focus is on the SVD of the resistance distance matrix obtained using the electrical structure of the power network. We do the SVD analysis of the bus admittance matrix as well, and shown that the coupling structure that results out of the admittance matrix is different from that obtained using the resistance distance matrix. Also, we do not explore control or communication architectures in this paper. Furthermore, unlike \cite{Dehghani2013}, our objective is complete network observability without zero injection measurements; there is no ranking of the PMUs to satisfy any constraints.

There are several other instances where the SVD has been invoked to study electric power systems. However, in the interest of space, we only mention references \cite{Tiranuchit1988} - \nocite{Osowski1994}\nocite{Hamdan1999}\nocite{Li2000}\nocite{Madtharad2003a}\nocite{Janik2009}\nocite{Taylor2012}\cite{Hernandez2013} without going into the details of the results presented therein. Though this list is by no means exhaustive, it highlights the wide spectrum of applications that can be realized using the SVD-based analysis in power engineering.
\vspace{-0.15cm}
\section{Optimal Set of PMUs}\label{sec:optimalset_pmu}
In this section, we first formulate the integer linear program to obtain the optimal or minimum number of PMUs. We then present the adjacency matrices for the topological and electrical structures-based PMU placement in \secref{subsec:pmuplacement_topology} and \secref{subsec:pmuplacement_electrical}, respectively. This section was also presented in \cite{Nagananda2013a}; however, it is reproduced here for sake of completeness.

Given a power network with $N$ buses and $M$ branches, we only consider complete network observability without conventional measurements. We let $\bm{A}$ denote the binary connectivity matrix of dimensions $N\times N$, $\bm{x}$ with dimensions $N\times 1$ denote the binary decision variable vector defined as follows:
\begin{eqnarray}\label{eq:binarydecision_vector}
x_i = \begin{cases}
1, ~\text{if a PMU is installed at bus}~i,\\
0, ~\text{otherwise},
\end{cases}
\end{eqnarray}
where $i$ = 1,\dots,$N$, and $\bm{b}$ is a unit vector of dimensions $N\times 1$. The PMU placement problem is formulated as follows:
\begin{eqnarray}\label{eq:pmuplacement}
\nonumber \min \sum_{i=1}^{N}x_i\\
\text{such that}~ \bm{A}\bm{x} \geq \bm{b}\\
\nonumber x_i \in \{0,1\}
\end{eqnarray}

\subsection{Topological structure-based PMU placement}\label{subsec:pmuplacement_topology}
In this case, the adjacency or connectivity matrix is derived directly from the bus admittance matrix. The entries of the bus admittance matrix are transformed into binary form, and used in the problem setup \eqref{eq:pmuplacement}. The entries of $\bm{A}$ are given by
\begin{eqnarray}\label{eq:topological_A}
\bm{A}:
\begin{cases}
a_{ij} = 1, ~\text{if}~i=j,\\
a_{ij} = 1, ~\text{if}~i~\text{and}~j~\text{are connected},\\
a_{ij} = 0, ~\text{if}~i~\text{and}~j~\text{are not connected}.
\end{cases}
\end{eqnarray}
Using the above described $\bm{A}$ in \eqref{eq:pmuplacement}, we obtain the topological structure-based PMU placement. Some of the above mentioned references let the location of $1$'s in the resulting optimal $\bm{x}$ to be the optimal location of PMUs, without fully utilizing the electrical properties of the network.
\vspace{-0.20cm}
\subsection{Electrical structure-based PMU placement}\label{subsec:pmuplacement_electrical}
In this subsection, the binary connectivity matrix is derived using the resistance distance between buses in the network. The resistance distance is the effective resistance between points in a network of resistors

Consider a network with $N$ nodes, described by the conductance matrix $\bm{G}$. Let $V_j$ and $g_{ij}$ denote the voltage magnitude at node $j$ and the conductance between nodes $i$ and $j$, respectively. The current injection at node $i$ is then given by
\begin{eqnarray}\label{eq:current_injection}
I_i = \sum_{j=1}^{N}g_{ij}V_j.
\end{eqnarray}
$\bm{G}$ acts as a Laplacian matrix to the network, provided there are no connections to the ground, {\ie}, if $\bm{G}$ has rank $N-1$. The singularity of $\bm{G}$ can be overcome by letting a node $r$ have $V_r = 0$. The conductance matrix associated with the remaining $N-1$ nodes is full-rank, and thus we have
\begin{eqnarray}\label{eq:nonreferencenodes}
\bm{V}_k = \bm{G}^{-1}_{kk}\bm{I}_k, k \neq r.
\end{eqnarray}
Let the diagonal elements of $\bm{G}^{-1}_{kk}$ be denoted $g^{-1}_{kk}$, $\forall k$, indicating the change in voltage due to current injection at node $k$ which is grounded at node $r$. The voltage difference between a pair of nodes $(i,j)$, $i\neq j\neq r$, is computed as follows:
\begin{eqnarray}\label{eq:voltage_difference}
e(i,j) = g^{-1}_{ii} + g^{-1}_{jj} - g^{-1}_{ij} - g^{-1}_{ji},
\end{eqnarray}
indicating the change in voltage due to injection of $1$ Ampere of current at node $i$ which is withdrawn at node $j$. $e(i,j)$ is called the resistance distance between nodes $i$ and $j$, and describes the sensitivity between current injections and voltage differences. In matrix form, letting $\boldsymbol{\Gamma} \triangleq \text{diag}(\bm{G}^{-1}_{kk})$, we have $\forall k \neq r$
\begin{eqnarray}
\bm{E}_{kk} &=& \boldsymbol{1}\boldsymbol{\Gamma}^{\mathrm{T}} + \boldsymbol{\Gamma}\boldsymbol{1}^{\mathrm{T}} - \bm{G}^{-1}_{kk} - \left[\bm{G}^{-1}_{kk}\right]^{\mathrm{T}},\label{eq:resistance_distance1}\\
\bm{E}_{rk} &=& \boldsymbol{\Gamma}^{\mathrm{T}},\label{eq:resistance_distance2}\\
\bm{E}_{kr} &=& \boldsymbol{\Gamma}.\label{eq:resistance_distance3}
\end{eqnarray}
The resistance distance matrix $\bm{E}$, thus defined, possesses the properties of a metric space \cite{Klein1993}.

To derive the sensitivities between power injections and phase angles, we start with the upper triangular part of the Jacobian matrix obtained from the power flow analysis, for the distance matrix to be real-valued:
\begin{eqnarray}\label{eq:upper_jacobian}
\Delta \bm{P} = \left[\frac{\partial P}{\partial \theta}\right]\Delta \theta + \left[\frac{\partial P}{\partial |V|}\right]\Delta |V|.
\end{eqnarray}
The matrix $\left[\frac{\partial P}{\partial \theta}\right]$ will be used to form the distance matrix, by assuming the voltages at the nodes to be held constant, {\ie}, $|V|=0$. It was observed that $\left[\frac{\partial P}{\partial \theta}\right]$ possesses most of the properties of a Laplacian matrix. By letting $\bm{G} = \left[\frac{\partial P}{\partial \theta}\right]$, the resulting distance matrix $\bm{E}$ measures the incremental change in phase angle difference between two nodes $i$ and $j$, $(\theta_i - \theta_j)$, given an incremental average power transaction between those nodes, assuming the voltage magnitudes are held constant. It was proved in \cite[Appendix]{Cotilla-Sanchez2012} that $\bm{E}$, thus defined, satisfies the properties of a distance matrix, as long as all series branch reactance are nonnegative.

For a power grid with $N$ buses, the distance matrix $\bm{E}$ translates into an undirected graph with $N(N-1)$ weighted branches. In order to compare the grid with an undirected network without weights, one has to retain the $N$ buses, but replace the $M$ branches with $M$ smallest entries in the upper or lower triangular part of $\bm{E}$. This results in a graph of size $\{N,M\}$ with edges representing electrical connectivity rather than direct physical connections. The adjacency matrix $\bm{B}$ of this graph is obtained by setting a threshold, $\tau$, adjusted to produce exactly $M$ branches in the network:
\begin{eqnarray}\label{eq:adjcency_matrix}
\bm{B}:
\begin{cases}
b_{ij} = 1, ~\forall e(i,j) < \tau,\\
b_{ij} = 0, ~\forall e(i,j) \geq \tau
\end{cases}
\end{eqnarray}
Letting $\bm{A} = \bm{B}$ in \eqref{eq:pmuplacement}, we obtain the electrical structure-based optimal PMU set. Next, we address the location problem.

\section{PMU placement}\label{sec:pmu_placement}
In this section, we perform the SVD-based analysis of the bus admittance and resistance distance matrices that helps us explore the coupling structure of the power network. The optimal PMU set obtained in the previous section will then be placed on the grid by exploiting this coupling structure.

Given an $N\times N$ matrix $\bm{P}$, the SVD of $\bm{P}$ is given by
\begin{eqnarray}\label{eq:svd}
\bm{S} = \bm{U}\boldsymbol{\Sigma}\bm{V}^{\ast},
\end{eqnarray}
where $\bm{U}$ is an $N\times N$ unitary matrix whose columns are the referred to as the left-singular vectors of $\bm{S}$, $\boldsymbol{\Sigma}$ is an $N\times N$ diagonal matrix whose diagonal entries are the singular values of $\bm{S}$, and $\bm{V}^{\ast}$ is an $N\times N$ unitary matrix whose columns are the referred to as the right-singular vectors of $\bm{S}$. $\bm{V}^{\ast}$ is the conjugate transpose of $\bm{V}$.

For a power network with $N$ buses, let $\bm{Y}$ be the $N\times N$ bus admittance matrix, while $\bm{E}$ is the $N\times N$ resistance distance defined by \eqref{eq:resistance_distance1} - \eqref{eq:resistance_distance3}. The SVDs of $\bm{Y}$ and $\bm{E}$ are given, respectively, by
\begin{eqnarray}
\bm{Y} &=& \bm{U}_{Y}\boldsymbol{\Sigma}_Y\bm{V}^{\ast}_Y,\label{eq:svd_Y}\\
\bm{E} &=& \bm{U}_{E}\boldsymbol{\Sigma}_E\bm{V}^{\ast}_E,\label{eq:svd_E}
\end{eqnarray}
where the notation is similar to that in \eqref{eq:svd}. The $N \times 1$ left and right singular vectors of the bus admittance matrix $\bm{Y}$ are denoted by $\bm{u}_{n,Y}$ and $\bm{v}_{n,Y}$, respectively, while those for the resistance distance matrix $\bm{E}$ are denoted by $\bm{u}_{n,E}$ and $\bm{v}_{n,E}$, respectively, where $n$ = 1,\dots,$N$ denotes the $n^{\text{th}}$ bus. The magnitude of the entries of these singular vectors signify the amount of electrical coupling that each bus shares with other buses. The diagonal elements of $\boldsymbol{\Sigma}_Y$ represent the singular values of $\bm{Y}$ and are denoted by $\sigma_{1,Y},\dots,\sigma_{N,Y}$, while the singular values of $\bm{E}$ are denoted by $\sigma_{1,E},\dots,\sigma_{N,E}$.

In the following, we provide a step-by-step procedure for PMU placement using the electrical structure-based approach.
\begin{enumerate}[1.]
\item The optimal or minimum number of PMUs is obtained by solving \eqref{eq:pmuplacement}. Let this number be denoted by $P(<N)$.
\item Compute the SVD of the resistance distance matrix to obtain the singular values and singular vectors.
\item Compute the magnitude of the vectors $\sigma_n\bm{u}_{n,E}$.
\item Pick $P$ vectors in the decreasing order of the value of the magnitudes of $\sigma_n\bm{u}_{n,E}$. These are labeled $\tilde{\bm{u}}_{p,E}$, $p=1,\dots,P$.
\item Lastly, a PMU is installed on the entry of $\tilde{\bm{u}}_{p,E}$ having the highest absolute value. Note that, the index of each entry of $\tilde{\bm{u}}_{p,E}$ corresponds to a bus number or location.
\end{enumerate}
In step 5, there is a possibility of conflict. For example, let the fourth entry have the highest absolute value in the singular vector $\tilde{\bm{u}}_{1,E}$ - this suggests that a PMU be installed at bus numbered four on the network. Now, let us suppose that the fourth entry in the singular vector $\tilde{\bm{u}}_{3,E}$ has the highest absolute value in that vector. This leads to a discrepancy, since a PMU was already placed on the fourth bus after computing the absolute values of the entries of $\tilde{\bm{u}}_{1,E}$. In such situations, the natural solution is to place the PMU on the entry of $\tilde{\bm{u}}_{3,E}$ with the second highest absolute value. This solution easily generalizes to all the $P$ vectors.

The conflict and its resolution can be summarized as follows: Let us consider two vectors $\tilde{\bm{u}}_{i,E}$ and $\tilde{\bm{u}}_{j,E}$, with $i\neq j$. Let the $k^{\text{th}}$ entry be the one with the largest absolute value in $\tilde{\bm{u}}_{i,E}$, and let the $\ell^{\text{th}}$ entry have the largest absolute value in $\tilde{\bm{u}}_{j,E}$. If $k=\ell$, then we give a higher priority to the vector having the larger magnitude. For instance, if $||\tilde{\bm{u}}_{i,E}|| > ||\tilde{\bm{u}}_{j,E}||$, we first place the PMU on the $k^{\text{th}}$ bus of $\tilde{\bm{u}}_{i,E}$. Then, we place a PMU on that entry of $\tilde{\bm{u}}_{j,E}$ having the next largest absolute value compared with its $\ell^{\text{th}}$ entry.

Thus, we have utilized the electrical structure and coupling of the power network to solve the PMU placement problem. For the topological structure-based PMU placement, the resistance distance matrix is replaced by the bus admittance matrix in the above devised procedure.
\vspace{-0.05cm}
\section{Results and discussion}\label{sec:results_discussion}
In this section, we first obtain the optimal set of PMUs for the IEEE test bus systems using the electrical and topological structure based approaches developed in this paper. In \secref{subsec:location_ambiguity}, we review the method devised in \cite[Section IV]{Nagananda2013a} to solve the location problem and bring out the ambiguity encountered in that approach. Finally, in \secref{subsec:coupling_location}, we show how the ambiguity can be resolved by exploiting the coupling structure of the network.

The bus and branch data, required to derive the bus admittance and resistance distance matrices, were obtained from archived resources \cite{Washington}. The binary integer programming tool of Matlab was used to solve the problem defined by \eqref{eq:pmuplacement}. We first tabulate, in \tabref{tab:bus_pmu}, the optimal number of PMUs obtained by solving the integer linear program \eqref{eq:pmuplacement}.
\begin{table}[h]
\centering
\begin{tabular}{ | c | c | c |}\hline
    IEEE bus system & Topological structure & Electrical structure \\ \hline
    9 & 3 &  4\\ \hline
    14 & 4 & 7 \\ \hline
    30 & 10 & 17 \\ \hline
    39 & 13 & 22 \\ \hline
    57 & 17 & 35 \\ \hline
    118 & 32 & 93 \\ \hline
    162 & 43 & 125 \\ \hline
    \end{tabular}
    \caption{Minimum number of PMUs based on topological and electrical structures for IEEE test bus systems}
    \label{tab:bus_pmu}
\end{table}
As shown in \tabref{tab:bus_pmu}, for each test bus system, the electrical structure-based approach results in a larger optimal set of PMUs is obtained compared with those resulting from the topology based approach for complete network observability without zero injection measurements. This is due to the fact that the electrical structure-based approach provides a more comprehensive description of the electrical connectedness between buses in the grid.
\vspace{-0.25cm}
\subsection{Approach of \cite{Nagananda2013a} to solve the location problem}\label{subsec:location_ambiguity}
In \cite[Section IV]{Nagananda2013a}, the entries of $\bm{B}$, given by \eqref{eq:adjcency_matrix}, were used to define the average resistance distance of each bus to other buses in the network:
\begin{eqnarray}\label{eq:average_distance}
\lambda_i = \sum_{j=1}^{N}\frac{b_{ij}}{N-1}.
\end{eqnarray}
We let
\begin{eqnarray}
\bm{\lambda} &\triangleq& \left[\lambda_1,\dots,\lambda_N\right],\label{eq:lambda_vector} \\
\lambda_{\min} &\triangleq& \min(\bm{\lambda}). \label{eq:lambda_minimum}
\end{eqnarray}
The central idea was that, if $\lambda_i > \lambda_{\min}$, a PMU need not be placed at the location of the $i^{\text{th}}$ bus. This was justified by the fact that, since $\lambda_i$ quantified the amount of average electrical connectivity between the $i^{\text{th}}$ and other buses in the network, the higher the value of $\lambda_i$ lower was the necessity to place a PMU on that bus. A pictorial representation provides a clearer picture.

\begin{figure}[h]
\centering
  \includegraphics[height=3in,width=3.5in]{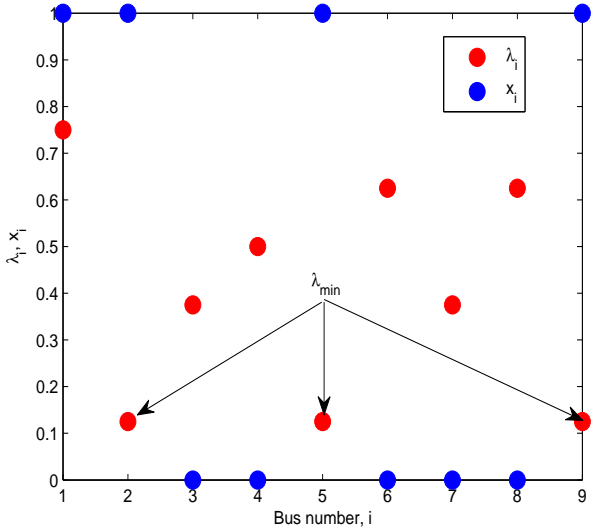}
  \caption{Average resistance distance for each bus for IEEE 9-bus system.}
  \label{fig:lambda_x_9bus}
\end{figure}
From the connectivity matrix $\bm{B}$ for the IEEE 9-bus system, and following \eqref{eq:average_distance}, a plot of $\lambda_i$ for each bus $i$ = 1,\dots,9 is shown in \figref{fig:lambda_x_9bus}. The binary decision variable vector $\bm{x}$, obtained by solving \eqref{eq:pmuplacement} for the 9-bus system is also shown in the same plot. It can be seen that, $x_i$ = 1 ({\ie}, PMU to be installed) only when $\lambda_i = \lambda_{\min}$; for all other values of $\lambda_i$, $x_i$ = 0 ({\ie}, no PMU). Therefore, we infer that the PMUs are to be installed on buses numbered 1, 2, 5 and 9. However, we see that $x_1$ = 1 though $\lambda_1 > \lambda_{\min}$. In \cite{Nagananda2013a}, the graphical structure of the adjacency matrix $\bm{B}$ was utilized to resolve this issue, and is discussed next.

\begin{figure}
\centering
  \includegraphics[height=3in,width=3.5in]{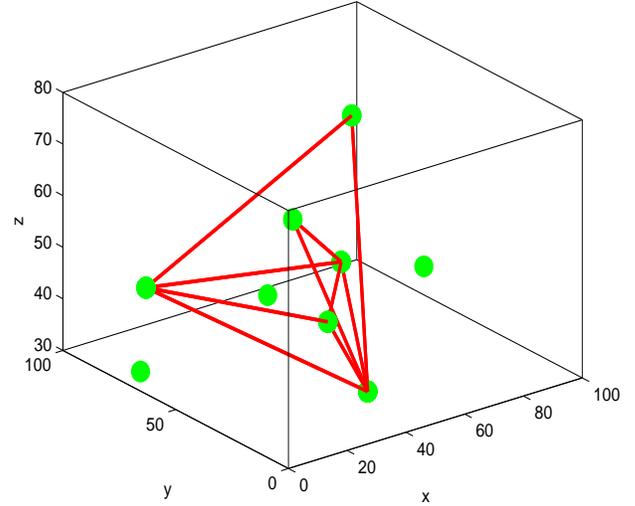}
  \caption{Graph corresponding to adjacency matrix $\bm{B}$ for IEEE 9-bus system.}
  \label{fig:elec_9bus}
\end{figure}
The graph corresponding to the adjacency matrix $\bm{B}$ for the IEEE 9-bus system is shown in \figref{fig:elec_9bus}. The disconnected points in the graph correspond to buses with very low electrical connectivity to remaining buses in the network, which suggests that a PMU must be installed at these buses, while a single PMU could be placed on an arbitrary node of the fully-connected subgraph to ensure complete network observability. Therefore, from \figref{fig:elec_9bus}, we see that the optimal number of PMUs required for the $9-$bus system is 3+ 1 =4, where 3 corresponds to the disconnected points, while 1 corresponds to the fully-connected subgraph.

To address the location problem, the following method was adopted. It is clear from the preceding discussion that the fully-connected subgraph shown in \figref{fig:elec_9bus} corresponds to the point at which $\lambda_i > \lambda_{\min}$ in \figref{fig:lambda_x_9bus}, since the connectivity is high between the nodes of this subgraph. Then, a PMU could be installed at any one of the points of this fully-connected subgraph. It was suggested in \cite{Nagananda2013a} that, without loss of generality, a PMU be installed at the bus numbered $1$ where $\lambda_1 > \lambda_{\min}$, which is in exact agreement with the analysis carried out using \figref{fig:lambda_x_9bus}. Therefore, the 4 PMUs were placed at buses numbered 1, 2, 5 and 9. However, one could argue to place the PMU on \emph{any point} of the fully-connected subgraph, {\ie}, not necessarily at the bus numbered 1, thereby leading to an ambiguity. The main goal of this paper is to resolve this ambiguity, and we accomplish this using the coupling structure of the grid.
\vspace{-0.25cm}
\subsection{The coupling structure to address the location problem}\label{subsec:coupling_location}
\begin{figure}[h]
\centering
  \includegraphics[height=3in,width=3.5in]{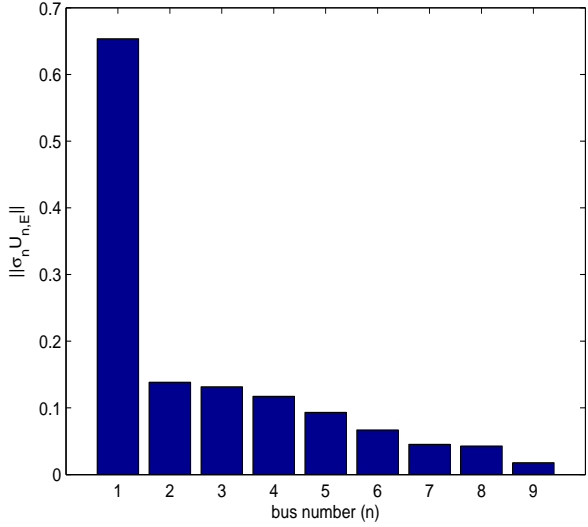}
  \caption{$||\sigma_n\bm{u}_{n,E}||$, $n=$ 1,\dots,9, for the IEEE 9-bus system.}
  \label{fig:elec_sigmaU_n_9bus}
\end{figure}

\begin{figure}[h]
\centering
  \includegraphics[height=3in,width=3.5in]{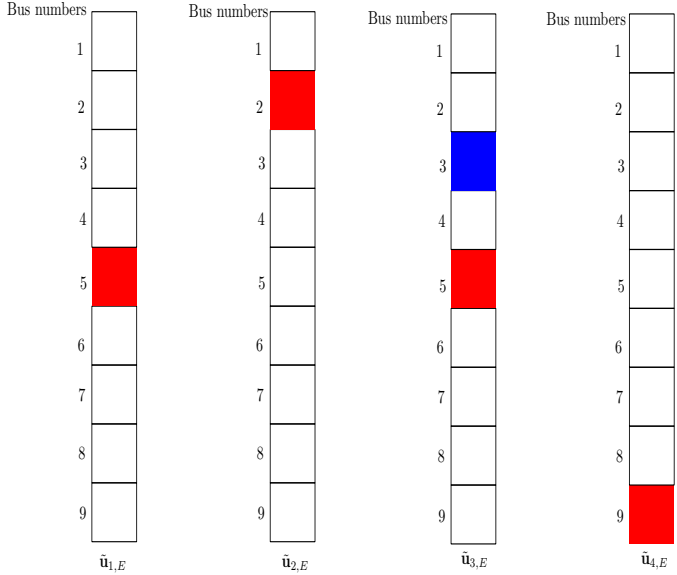}
  \caption{Pictorial representation of the vectors $\tilde{\bm{u}}_{p,E}$, $p$ = 1,\dots,4 for the IEEE 9-bus system.}
  \label{fig:elecplace9bus}
\end{figure}

In this subsection, we follow the procedure developed in \secref{sec:pmu_placement} to place the optimal set of PMUs obtained from \secref{sec:optimalset_pmu}. We again consider the IEEE 9-bus system for our analysis, which required an optimal number of 4 PMUs (see \tabref{tab:bus_pmu}) to be installed on the network. We first compute the SVD of the resistance distance matrix ($\bm{E}$) to get the singular values and singular vectors. The plot of the magnitude of $\sigma_n\bm{u}_{n,E}$, $n=$ 1,\dots,9, is shown in \figref{fig:elec_sigmaU_n_9bus}.
We pick $P=$ 4 vectors in the decreasing order of the values of the magnitude of the vectors $\sigma_n\bm{u}_{n,E}$ (the four leftmost vectors in \figref{fig:elec_sigmaU_n_9bus}) and label them $\tilde{\bm{u}}_{p,E}$, $p$ = 1,\dots,4. We represent the vectors $\tilde{\bm{u}}_{p,E}$ as shown in \figref{fig:elecplace9bus}, where a column denotes a vector, while a box in each column denotes an entry of the vector. The number of boxes in each column equals the number of buses, with the topmost box denoting the first bus and the bottommost the last bus in the network.

In each column, the entry with the largest absolute value is marked in red, and a PMU is placed on the bus corresponding to this entry. Accordingly, a PMU is placed on the 5$^{\text{th}}$ bus corresponding to $\tilde{\bm{u}}_{1,E}$ and on the 2$^{\text{nd}}$ bus corresponding to $\tilde{\bm{u}}_{2,E}$. As pointed out in \secref{sec:pmu_placement}, there arises a conflict: in both the vectors $\tilde{\bm{u}}_{1,E}$ and $\tilde{\bm{u}}_{3,E}$, the entry having the largest absolute value appears in the 5$^{\text{th}}$ location. However, since a PMU was already placed on the 5$^{\text{th}}$ bus corresponding to $\tilde{\bm{u}}_{1,E}$, for $\tilde{\bm{u}}_{3,E}$ we place a PMU on the entry having the second largest absolute value. Therefore, as shown in \figref{fig:elecplace9bus}, a PMU is placed on the 3$^{\text{rd}}$ entry (marked in blue), which has the second largest absolute value in $\tilde{\bm{u}}_{3,E}$. Finally, the fourth PMU is placed on the 9$^{\text{th}}$ bus corresponding to the entry with the largest absolute value in $\tilde{\bm{u}}_{4,E}$.

To summarize, for the IEEE 9-bus system, using the electrical structure-based approach to PMU placement, the optimal number of PMUs obtained was 4 and these PMUs are to be installed at buses numbered 2, 3, 5 and 9. Therefore, the decision to place the PMUs on buses numbered 2, 5 and 9 is in unison with the the results obtained in the previous subsection, while placing a PMU on the bus numbered 3 resolves the ambiguity encountered when using the method developed in \cite{Nagananda2013a} (where the PMU was placed on the bus numbered 1).

\begin{figure}[h]
\centering
  \includegraphics[height=3in,width=3.5in]{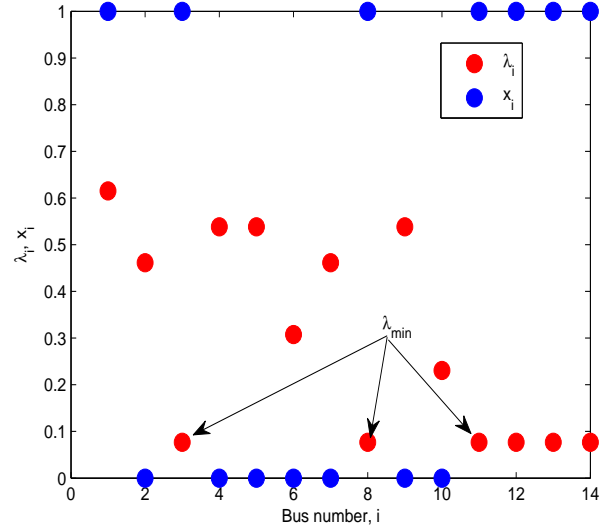}
  \caption{Average resistance distance for each bus for IEEE 14-bus system.}
  \label{fig:lambda_x_14bus}
\end{figure}
\begin{figure}[h]
\centering
  \includegraphics[height=3in,width=3.5in]{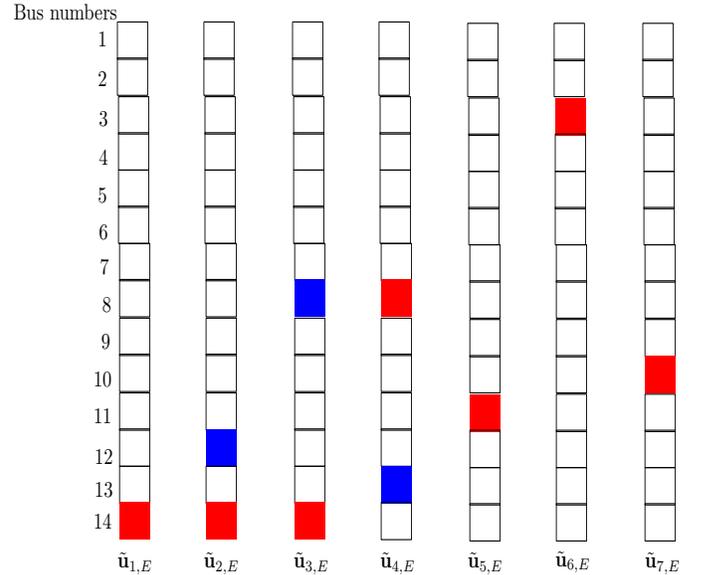}
  \caption{Pictorial representation of the vectors $\tilde{\bm{u}}_{p,E}$, $p$ = 1,\dots,7 for the IEEE 14-bus system.}
  \label{fig:elecplace14bus}
\end{figure}
For the IEEE 14-bus system, the electrical structure-based approach yields an optimal of 7 PMUs to be installed in the network. Following the method presented in \cite{Nagananda2013a}, six of these PMUs are located on buses numbered 3, 8, 11, 12, 13 and 14. Owing to the ambiguity, the seventh PMU will be installed on the bus numbered 1. On the other hand, analyzing the coupling structure of the 14-bus system leads us to install the seven PMUs on buses numbered 3, 8, 10, 11, 12, 13 and 14. Installing a PMU on the bus numbered 10, instead of on the bus numbered 1 (which was ambiguous), demonstrates the benefit of the method developed in this paper. As before, for ease of understanding, we show the plots of the average resistance distance in \figref{fig:lambda_x_14bus} and the vectors $\tilde{\bm{u}}_{p,E}$, $p$ = 1,\dots,7, in \figref{fig:elecplace14bus}.

\begin{figure}[h]
\centering
  \includegraphics[height=3in,width=3.5in]{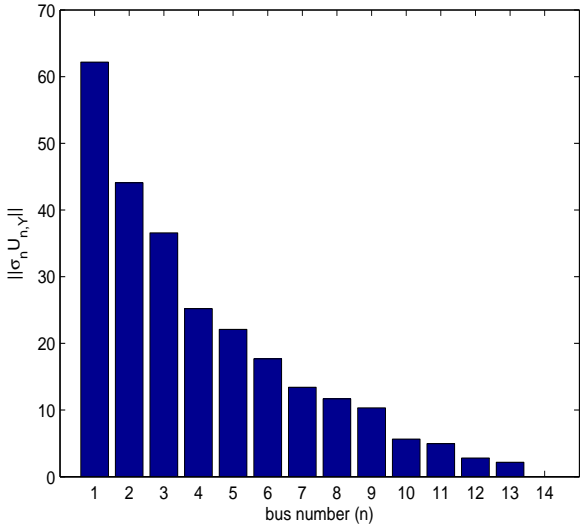}
  \caption{$||\sigma_n\bm{u}_{n,Y}||$, $n$ = 1,\dots,14, for the IEEE 14-bus system.}
  \label{fig:topo_sigmaU_n_14bus}
\end{figure}

\begin{figure}[h]
\centering
  \includegraphics[height=3in,width=3.2in]{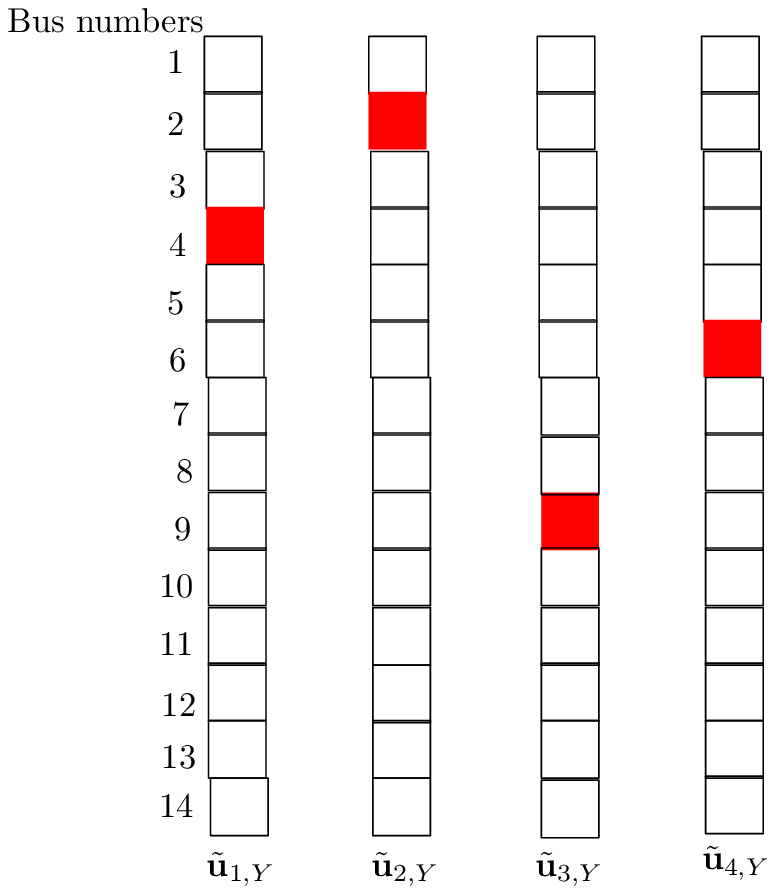}
  \caption{Pictorial representation of the vectors $\tilde{\bm{u}}_{p,Y}$, $p$ = 1,\dots,4, for the IEEE 14-bus system.}
  \label{fig:topoplace14bus}
\end{figure}

Similar analysis was carried out for the IEEE 14-bus system using the topological structure, where the resistance distance matrix is replaced by the bus admittance matrix of the power network. From \tabref{tab:bus_pmu}, we see that the minimum number of PMUs required for this approach is 4. Employing the procedure specified in \secref{sec:pmu_placement}, we infer that the optimal locations to install these 4 PMUs are at buses numbered 2, 4, 6 and 9. The plot of the magnitudes of the vectors $\sigma_n\bm{u}_{n,Y}$, $n$ = 1,\dots,14, shown in  \figref{fig:topo_sigmaU_n_14bus} and the pictorial representation of the vectors $\tilde{\bm{u}}_{p,Y}$, $p$ = 1,\dots,4, shown in \figref{fig:topoplace14bus} are used to obtain these results.

\begin{figure}[h]
\centering
  \includegraphics[height=3in,width=3.5in]{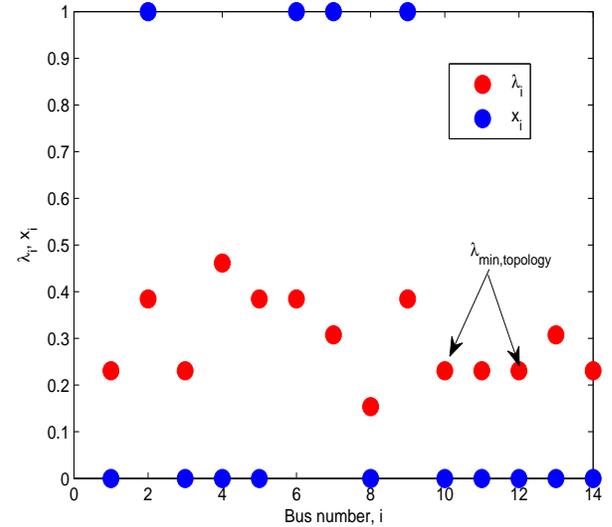}
  \caption{Average topological distance for each bus for IEEE 14-bus system.}
  \label{fig:lambda_x_14bus_topo}
\end{figure}
Now, let us compare the above results with those obtained using the method presented in \cite{Nagananda2013a}. Similar to the average resistance distance, we first define the average topological distance as follows:
\begin{eqnarray}\label{eq:average_distance_topo}
\lambda_{i,\text{topological}} = \sum_{j=1}^{N}\frac{a_{ij}}{N-1},
\end{eqnarray}
where $a_{ij}$'s are specified by \eqref{eq:topological_A}. A plot of the average topological distance for each bus is shown in \figref{fig:lambda_x_14bus_topo}, which suggests that the four PMUs are to be placed on buses numbered 2, 6, 7 and 9. This is not consistent with the results obtained using the electrical structure of the grid. Furthermore, the plot of the average topological distance for each bus shown in \figref{fig:lambda_x_14bus_topo} does not exhibit a pattern, unlike the average resistance distance. That is, $x_i$ does not equal 1 even when $\lambda_i = \lambda_{\min,\text{topology}}$. Similar inconsistencies were experienced for other IEEE bus systems. However, owing to space limitation, we do not present the results for other bus systems here.

From the above analysis we infer that the topological structure-based approach to PMU placement suffers from inconsistencies. Whereas, the electrical structure-based approach provides a more reliable framework. Thus, we conclude that the electrical structure-based approach provides a comprehensive framework for PMU placement and, given the significance of the problem, this approach should be frontrunner for PMU placement in electric power systems.

In the process of solving the location problem, we have uncovered connections between the average resistance distance $\lambda_i$, given by \eqref{eq:average_distance}, and the vectors $\tilde{\bm{u}}_{p,E}$, $p$ = 1,\dots,$P$. We noticed that when $\lambda_i = \lambda_{\min}$, a PMU was installed at the $i^{\text{th}}$ bus. On the other hand, this $i$ also corresponded to an entry in the vector $\tilde{\bm{u}}_{p,E}$ having a large absolute value; it may be the largest, or the $n^{\text{th}}$ largest depending on the ``conflict-resolution'' technique presented in \secref{sec:pmu_placement}. This hints at the existence of a functional relationship between the adjacency matrix given by \eqref{eq:adjcency_matrix} and the singular vectors $\tilde{\bm{u}}_{p,E}$, $p$ = 1,\dots,$P$. A complete characterization of this function is of fundamental importance and is an important open problem. Interestingly, we do not observe this kind of a relationship between the average topological distance given by \eqref{eq:average_distance_topo} and the vectors $\tilde{\bm{u}}_{p,Y}$, $p$ = 1,\dots,$P$. This can be attributed to the fact that the topological structure does not provide a comprehensive description of the electrical connectivity between the nodes in the network.

\section{Conclusion}\label{sec:conclusion}
The SVD-based analysis of the resistance distance matrix revealed the electrical coupling between the components of the network. The coupling was utilized to solve the location problem in a refined manner, thereby resolving the ambiguity encountered in \cite{Nagananda2013a}, where the graphical structure of the adjacency matrix derived using the resistance distance was employed to solve this problem. A potential drawback of the method proposed in this paper is that the SVD might lead to a computational bottleneck for power networks with large number of buses. However, with developments in fast methods for large scale SVD (for {\eg}, see \cite{Woolfe2008}), the electrical structure-based approach to PMU placement holds promise. From an operational point of view, like network observability, stability, state estimation, {\etc}. where PMU measurements are being included, our approach is more appealing since it provides a comprehensive description of the electrical connectivity between network components. Characterizing the functional relationship between the adjacency matrix obtained using the electrical structure of the network and the singular vectors of the resistance distance matrix is a fundamental open problem, with serious implications to related areas.

\bibliographystyle{IEEEtran}
\bibliography{IEEEabrv,powersystems}

\end{document}